\documentclass[]{aa}
\usepackage{graphicx}
\usepackage{psfig}
\newcommand{\magunit}{mag arcsec$^{-2}$}
\newcommand{\densunit}{M$_{\sun}\ \mbox{pc}^{-2}$}
\newcommand{\sfrdensunit}{M$_{\sun}\ \mbox{yr}^{-1} \mbox{kpc}^{-2}$}
\newcommand{\sfrunit}{M$_{\sun}\ \mbox{yr}^{-1}$}

\newcommand{\smal}{half-major axis length}
\newcommand{\sfrdix}{SFR$_\mathrm{10}$}
\newcommand{\sfrcent}{SFR$_\mathrm{100}$}
\newcommand{\barZ}{$\bar{Z}$}
\newcommand{\ltot}{L$_\mathrm{tot}$}
\newcommand{\ltotabs}{L$^\mathrm{abs}_\mathrm{tot}$}
\newcommand{\ml}{mass-to-light ratio}
\newcommand{\dB}{R$^\mathrm{B}_\mathrm{25}$}
\newcommand{\dR}{R$^\mathrm{R}_\mathrm{23.5}$}
\newcommand{\dK}{R$^\mathrm{K}_\mathrm{21.5}$}
\newcommand{\dH}{R$^\mathrm{H}_\mathrm{20.5}$}

\newcommand{\ctrois}{{\tt C}$_3$}
\begin{document}

\title{Photometric and dynamic evolution of an isolated disc galaxy
simulation}

\author{L\'eo Michel--Dansac\inst{1} \and\ Herv\'e Wozniak\inst{1,2}}

\offprints{L\'eo Michel--Dansac, \email{leo.michel-dansac@oamp.fr}}

\institute{Observatoire Astronomique de Marseille-Provence, Laboratoire
  d'Astrophysique de Marseille, 2 Place Le Verrier, 13248 Marseille
  Cedex 4, France \and \emph{Present address:} Centre de Recherche
  Astronomique de Lyon, 9 av. Charles Andr\'e, 69561 Saint-Genis Laval
  cedex, France}

\date{Received / Accepted}

\authorrunning{Michel--Dansac \& Wozniak}
\titlerunning{Photometric and dynamic evolution of an isolated disc galaxy
simulation}{}

\abstract{We present a detailed analysis of the evolution of a
simulated isolated disc galaxy. The simulation includes stars, gas,
star formation and simple chemical yields. Stellar particles are
split in two populations: the old one is present at the beginning
of the simulation and is calibrated according to various ages and
metallicities; the new population borns in the course of the
simulation and inherits the metallicity of the gas particles.  The
results have been calibrated in four wavebands with the
spectro-photometric evolutionary model GISSEL2000 (Bruzual \&\
Charlot~\cite{BC93}). Dust extinction has also been taken into
account.  A rest-frame morphological and bidimensional photometric
analysis has been performed on simulated images, with the same tools
as for observations.  The effects of the stellar bar formation and the
linked star formation episode on the global properties of the galaxy
(mass and luminosity distribution, colours, isophotal radii) have been
analysed. In particular, we have disentangled the effects of stellar
evolution from dynamic evolution to explain the cause of the isophotal
radii variations.  We show that the dynamic properties (e.g. mass) of
the area enclosed by any isophotal radius depends on the waveband and
on the level of star formation activity.  It is also shown that the
bar isophotes remain thinner than mass isodensities a long time ($>
0.7$~Gyr) after the maximum of star formation rate. We show that bar
ellipticity is very wavelength dependent as suggested by real
observations.  Effects of dust extinction on photometric and
morphological measurements are systematically quantified. For
instance, it is shown that, when the star formation
rate is maximum, no more than 20\%\ of the B band luminosity can escape
from the bar region whereas, without dust extinction, bar B band luminosity
accounts for 80\%\ of the total B band luminosity. Moreover, the extinction
is not uniformly distributed inside the bar.

}

\maketitle
\section{Introduction}

Early N-body models of isolated disc galaxies, using only
collisionless particles, have clarified how several morphological
features such as bars, spiral arms (e.g. Hohl \cite{H78}) and boxy
bulges (Combes \& Sanders \cite{CS81}), appear in stellar discs and
evolve. Soon afterwards, it was realised that the role of the gas, the
dissipative component, cannot be neglected. Collisionless N-body codes
were thus coupled with hydrodynamic codes, either for cosmological
purposes (Katz \& Gunn \cite{KG91}), or for detailed studies of
isolated (Friedli \& Benz \cite{FB93}) or merging galaxies (Mihos \&
Hernquist \cite{MH96}). For isolated disc galaxies, this kind of codes
has brought clear evidence that the presence of gas can deeply change
the morphology on a short timescale (less than 1~Gyr).  For instance,
the gas can be responsible for the bar within bar phenomenon (Friedli
\& Martinet \cite{FM93}) or the dissolution of the bar and the
formation of a bulge by secular evolution (Pfenniger \& Norman
\cite{PN90}, Norman et al. \cite{NSH96}).  Studies of the fueling
mechanisms at work in central regions (e.g. Friedli \& Benz
\cite{FB93}, Shlosman \& Noguchi \cite{SN93}) have been also studied
with such hybrid codes.

However, stars and gas are not only bound together by gravitation, but
also by the process of star formation and the energy and chemical
feedback of supernov\ae\ and stellar winds. Since star formation is
not well understood even in our own Galaxy, several recipes have been
implemented to mimic star formation in hydro+stellar dynamic
codes. These recipes have allowed to study for the first time the
influence of bars on galactic abundance gradient (Friedli et
al. \cite{FBK94}), the impact of the stellar ultraviolet radiation on
star formation (Gerritsen \& Icke \cite{GI97}), the nuclear activity
(Heller \& Shlosman \cite{HS94}), the starbursts induced by mergers
(e.g. Mihos \& Hernquist \cite{MH96}), the renewal of bars by gas
accretion (Bournaud \& Combes \cite{BC02}), etc. They also allow to
perform self-consistent simulations of the formation and evolution of
galaxies (e.g. Katz \cite{K92}, Steinmetz \& Muller \cite{SM95}).

Whatever the kind of code used for the study of secular evolution, the
morphological analysis of the simulations focuses on the properties of
the mass distribution. To compare in details these models with
observations it is implicitly assumed that all particles have the same
mass-to-light ratio taken to unity for convenience. However, the real
galaxies are built of composite stellar populations, then of various
mass-to-light ratios and metallicities. Thus, in order to make more
straightforward comparisons of simulations with multi-wavelength
observations, we have used stellar population synthesis models to
photometrically calibrate our self-consistent simulations which
include stars, gas and star formation. We also take into account the
effect of dust extinction.

This approach has been recently used to study the formation and the
subsequent evolution of elliptical (e.g. Kawata \cite{K01}) or disc
galaxies in a $\Lambda$ cold dark matter scenario (e.g. Westera et
al. \cite{WSBG02}, Abadi et al. \cite{ANSE03}).  Our own approach
slightly differs from previous ones since we do not simulate the
formation of a disc galaxy in a primeval Universe. Indeed, we restrict
ourselves to make a detailed study of the short-term evolution of an
isolated disc galaxy already formed. All the particles are thus used
to simulate the evolution of the galaxy.

We describe in Sect.~2 the numerical model and our technique of
photometric calibration. The global evolution (i.e. integrated
properties) of the model is presented in Sect.~3, and the
morphological evolution in Sect.~4. We summarise our findings in
Sect.~5.

\section{The N-body simulation and photometric calibration}

\subsection{Description of the code}
We used {\tt PMSPHSF}, the N-body code developed in Geneva. It
includes stars, gas and recipes to simulate star formation. The broad
outlines of the code are the following: the gravitational forces are
computed with a particle--mesh method using a 3D polar grid with
$(N_R, N_\phi, N_Z)=(31,32,64)$ active cells (Pfenniger \& Friedli
\cite{PF93}), the hydrodynamics equations are solved using the SPH
technique (Friedli \& Benz \cite{FB93}).  Star formation recipes are
those implemented by Friedli \& Benz (\cite{FB95}). Here we briefly
summarise the main features. For a detailed description of this
modelling and a discussion on the influence of the free parameters, we
refer the reader to Friedli \& Benz (\cite{FB95}).

Star formation and feedback modelling is basically based on the
instantaneous star formation approximation. It consists in the
following steps~: the identification of the regions of star formation,
the conversion of a fraction of gas into stars and the computation of
the amount of energy and heavy elements injected in the interstellar
medium (i.e. the energy and chemical feedback). Because the latter
step leads to the heating of gas, a simple treatment of radiative
cooling is implemented into the energy equation.

The first task is to identify the gaseous particles that will form
stars. Friedli \& Benz (\cite{FB93}) examined some possible
criteria. Not surprisingly, they found that the standard Jeans
instability criterion can be applied to spherical non-rotating gaseous
systems, but that for rotating flat discs Toomre's instability
criterion (Toomre \cite{T64}) is a better indicator. Observational
evidence also appear to support the use of this criterion as a good
indicator for locating star formation at intermediate scale-lengths
(Kennicutt \cite{K89,K90}).

In any case, a particle $i$ will be assumed to undergo a
star formation episode if the following condition is verified:
$$
Q_{i}^{g} = \frac{s_{i}\kappa_{i}}{\pi G \Sigma_{i}^{g}} \leq \lambda
$$
where $Q_i^g$ is Toomre's parameter, $s_{i}$ is the local sound speed,
$\kappa_{i}$ is the generalised epicyclic frequency, $\Sigma_{i}^{g}$
is the gas surface density.  This condition is clearly Toomre's
stability criterion.  The constant $\lambda$ equals unity in the case
of an axisymmetric gaseous disc subject to radial instabilities
(cf. Toomre \cite{T64}). However, a value of $\lambda \approx 1.4$
(with $s = 6$ km s$^{-1}$) is derived from observations (Kennicutt
\cite{K90}).  

Once a particle of mass $m_{i}^{g}$ subject to star formation has been
identified, a fraction of its mass is converted into a stellar
particle while total energy and momentum are conserved. The fraction
of the gas particle actually transformed (i.e. the star formation
efficiency factor $\epsilon$) is a free parameter. Note that within
the framework of \emph{instantaneous} star formation, $\epsilon$ is
the product between the fraction of gas actually transformed into star
$\eta$ (in accord with the IMF) and the mass fraction $1-R$ finally
blocked in low mass stars or stellar remnants. Thus, we can write
$\epsilon = (1-R)\eta$. We re-use values of Friedli \& Benz
(\cite{FB95}) which are $R \approx 0.45$ (Maeder \cite{M92})
and $0.02 \leq \eta \leq 0.5$. This translates into
a possible range $0.01 \leq \epsilon \leq 0.25$. For the present work,
$\epsilon = 0.1$.

In practice, a particle $i$ subject to star formation is split
into one star particle $m_{k}^{\star,new}$ and one gas particle
$m_{i}^{g,new}$ of reduced mass. Mass conservation requires that
these masses be given by
$$
    m_{k}^{\star,new} = \epsilon m_{i}^{g}, \qquad m_{i}^{g,new} =
    (1-\epsilon) m_{i}^{g}.
$$ 
At this stage, note that no Schmidt law (Schmidt \cite{S59}) has
been imposed; the star formation rate (SFR) is thus a consequence of
the (hydro)dynamic instabilities (see Sect. 5.2 of Friedli \& Benz
\cite{FB95} for a discussion on the relation between the SFR and the
gas surface density).  The system can make several generations of
stars from one gas particle and can thus quickly adjust itself to
stabilise any collapsing region. However, in order to avoid numerical
problems associated with the existence of gas particles differing
widely in mass we restrict the allowable gas particle mass range to a
maximum of 20. For $\epsilon=0.1$, this corresponds to a maximum of 29
generations of new stars from a single gas particle.  Gas particles
which reach the lower mass limit are transformed into a single
collisionless particle.

Probably the most important effect of star formation on the global
evolution of a galaxy is connected to the large amount of energy
released in supernova explosions and stellar winds. In this modelling
of feedback, the contribution of \textsc{SNI}a supernov\ae\ is
neglected. Let $E_{i,tot}^{SN}$ be the total energy produced by
\textsc{SNII} explosions at particle $i$ at a given timestep
$$ 
E_{i,tot}^{SN} = E_{i,th}^{SN} + E_{i,mec}^{SN} = R^{SN} \bar{E}^{SN}
    (1-R)^{-1} m_{k}^{\star,new},
$$ 
where $E_{i,th}^{SN}=(1-f)E_{i,tot}^{SN}$ and
$E_{i,mec}^{SN}=fE_{i,tot}^{SN}$ are respectively the thermal and
mechanical energy injected, $R^{SN}$ is the number of SN created par
unit mass of stars formed, $\bar{E}^{SN}\approx10^{51}$ erg is the
mean energy produced by one \textsc{SNII}, and $f$ is a very poorly
determined free parameter between 0 and 1. It can be seen as the
transfer efficiency of mechanical energy by thermalization to the
surrounding ISM. For the present work, we use $f=0.01$ and $R^{SN}
=0.004$.

Depending on the model assumptions ($Z$, IMF, SFR), the asymptotic
value of energy ratio $R^{W}$ between hot winds and \textsc{SNII}
contributions can vary from almost 0 to about 2 (Leitherer et
al. \cite{Letal92}). In particular, an increase of metals strongly
increases $R^{W}$. So, the contribution of winds to the energy
injection is simply assumed to be proportional to the contribution
coming from SNe but with a linear metallicity dependence, i.e
$E_{i,tot}^{W}=R^{W}E_{i,tot}^{SN}$ with $R^{W}=25Z$.

Energy, in both forms, is injected instantaneously and is spread over
the $n_{i}$ neighbour particles $j$ of particle $i$ using 
the SPH smoothing kernel as the normalised
weighting function $W_{ij}$. The fraction of
mechanical energy given to particle $j$,
$W_{ij}(E_{i,mec}^{SN}+E_{i,mec}^{W})$, is added by computing the
corresponding velocity vector increment
$$
(\Delta \vec v)_{j}^{SN} = \sqrt{\frac{2W_{ij} f
 (1+R^{W})E_{i,tot}^{SN}}{m_{j}}}
 \frac{\vec{r}_{j}-\vec{r}_{i}}{|\vec{r}_{j}-\vec{r}_{i}|},\ 
 j=1,...,n_{i}.
$$
Similarly, the specific thermal energy $E_{i,th}^{SN}$ increment
received by particle $j$ is given by
$$
\Delta u_{j}^{SN}=\frac{W_{ij}(1-f)(1+R^{W})E_{i,tot}^{SN}}{m_{j}},\qquad
j=1,...,n_{i}.
$$

Since the instantaneous star formation approximation is used,
instantaneous recycling is assumed as well. Therefore, during
evolution, gas particles will see their abundances modified as star
formation proceeds in either immediate neighbourhood, whereas star
particles will be endowed at birth with a relative abundance pattern
corresponding to their parents gas cloud. Changes of star abundance
due to their evolution are neglected. Formally, the abundance change
of a gas particle over one timestep can be written as
$$
    Z_{j}^{g}(t+\Delta t)=
    Z_{j}^{g}(t)+\frac{y_{Z}m_{k}^{\star,new}W_{ij}}{m_{j}^{g}},
    \qquad j=1...n_{i}, i,
$$
where $W_{ij}$ is the weighting factor described previously, and
$y_{Z}$ is the net yield for metals (i.e. the mass of metals ejected
by all stars per unit mass locked into stars). The net yield are taken
from Maeder (\cite{M92}, model C, Table~7). At birth, star particles
are created from gas, thus they have the same chemical composition.

The method of treating radiation losses is to use tabulated cooling
functions from published data and to interpolate between them. For the
present work, we use those of Boehringer \& Hensler (\cite{BH89}) They
cover the temperature range $10^{4}$ K to $10^{8}$ K. Therefore, we
assume a lower cut-off of $T_{cut}=10^{4}$ K. Since we are not in a
position to compute accurately the evolution of all elements, we
restrict ourselves to cooling functions assuming solar abundance.

\subsection{Initial conditions}
\label{ssec:ci}
An initial stellar population is setup to reproduce a disc galaxy with
an already formed bulge. These particles form what we call hereafter
the `old population' as opposed to particles created during the
evolution (`new population').

The initial stellar positions for 500\,000 particles are drawn from a
superposition of two axisymmetric Miyamoto-Nagai discs of mass
respectively $10^{10}$ and $10^{11}$~M$_{\sun}$, of scale lengths
respectively 1.5 and 4~kpc and common scale height of 0.5~kpc. The
initial disc radius is 25~kpc.

Stellar velocities and velocity dispersions are found by solving
numerically the Jeans equations directly on the grid using an
iterative process. The initial conditions represent an equilibrium
solution of the stellar hydrodynamics equations, incorporating the
softened gravitation as well as the natural anisotropic velocity
dispersion tensor of disc. This obviously does not mean that the
stellar disc is fully stable. Indeed, the disc is dynamically cold
enough to be bar mode unstable.

The gaseous component is represented by 50\,000 particles drawn from a
third Miyamoto-Nagai disc of scale length of 6~kpc, scale height of
0.1~kpc and truncation radius of 25~kpc. The total gas mass amounts to
$1.1\,10^{10}$~M$_{\sun}$.  An initial gradient of metallicity has
been imposed following the relationship $ Z = 0.02 \times\exp(-r/r_Z)$
where $r_Z$, the scalelength of the metallicity distribution, is
4.34~kpc. There is no vertical gradient as the gaseous component is
initially distributed in a thin layer (scale height of 0.1~kpc).

We also ran the same simulation with a factor of 5 less particles to
check whether the SFR and its variation in time depends on the number
of particles. We find that our results reported in
Sect.~\ref{ssec:sfr} are robust although the details of the dynamic
evolution are obviously different.

For this first paper, we did not include any dark halo to be able in
the near future to compare the effects of having added such an
additional component. These initial conditions are identical to the
{\tt Run A} of Wozniak et al. (\cite{Wetal03}).

After 1~Gyr, the simulation numbers 1\,264\,729 particles, of which
26\,573 gas particles. In comparison, the stellar disc and the dark
matter halo of the final galaxy of Abadi et al. (\cite{ANSE03}) is
described by less than 166\,000 particles. The final galaxy of Samland
\& Gerhart (\cite{SG03}) contains 614\,500 stellar particles at the
end of the simulation embedded in a growing halo.

In Table~\ref{tab:param} we give the masses of the various components
(old and new populations, gas) for the initial time and after an
evolution of 1~Gyr.

\subsection{Stellar mass-to-light ratio calibration}
The output of any N-body simulation is mainly a set of positions and
velocities for all particles. In general, particles have the same
masses so that it is not necessary to make any assumption on their
mass-to-light ratio.  Thus, the problem of luminosity calibration has
been widely overlooked.  However, because of star formation, the
particles created during the run have various masses, ages and
metallicities. The observable properties, which are mean quantities
integrated along the line-of-sight (LOS) and over a spatial resolution
element, must be weighted by the {\em luminosity} of the particles.
Therefore, we need an accurate mass-to-light ratio to convert masses
into luminosities in various photometric bands.

For each stellar particle, given its age and metallicity, the
mass-to-light ratio $\Upsilon$ was obtained from a bi-linear
interpolation into the tables of GISSEL2000 (Bruzual \& Charlot
\cite{BC93}), for a Salpeter initial mass function (IMF), with mass
cut-off at 0.1 and 100 M$_{\sun}$.  Then, luminosity of the particle
$i$ (${L}^{i}$), in the waveband $X$, is computed according to its
mass (${M}^{i}$), birth metallicity ($Z_{\mathrm{born}}^{i}$) and the
time elapsed since its birth ($t-t_{\mathrm{born}}^{i}$):
$${L}^{i}_{{X}}(t) = \frac{{M}^{i}}
{\Upsilon_{{X}}(t-t^{i}_{\mathrm{born}},Z^{i}_{\mathrm{born}})}.
$$
It is noteworthy that the photometric calibration is not fully
consistent with star formation and feedback modelling since the
selected value of $R^{SN}$ corresponds to a lower mass limit for
\textsc{SNII} equal to $\sim 12$ M$_{\sun}$ whereas the standard value
8 M$_{\sun}$ is used in GISSEL2000.

For the old population (present at the beginning of the simulation) we
have assumed ages of 6.7 and 10.4~Gyr and metallicities of $Z=0.004$
and $Z=0.02$ (cf. Table~\ref{tab:models}). Obviously, this implicitly
assumes that all the old stars simultaneously born 6.7 or 10.4~Gyr
before the beginning of the simulation.  Assuming an observer located
at $z=0$ at the end of the simulation ($t=1$~Gyr), these calibrations
imply a redshift of formation of $z=1$ and $z=3$ for resp. models C-D
and models A-B.

\begin{table}[ht]
\caption[ ]{Calibration of the old population.}
\begin{center}
\begin{tabular}{lcc}
\hline
\hline
     & Age & Metallicity\\
\hline
Model A  & $10.4$ & $0.004$ \\
Model B  & $10.4$ & $0.020$ \\
Model C  & $6.7$ & $0.004$ \\
Model D  & $6.7$ & $0.020$ \\
\hline
\end{tabular}
\end{center}
\label{tab:models}
\end{table}

Global quantities (e.g. luminosities, mass-to-light ratios, etc.) and
surface brightness distribution were computed in various colours.  We
choose to work with four widely used broad band filters to cover the
wavelength domain from the visible to the near infrared, namely B, R,
H, and K in the Johnson's system of GISSEL2000. 

\subsection{Dust extinction}
Dust extinction in B, R, H and K band is simulated in the simplest
way, assuming a constant gas-to-dust ratio
$$ N(\ion{H}{i})/A_\mathrm{V} = 5.34\times 10^{21}\mbox{cm}^{-2}.$$
$A_\mathrm{V}$ is then converted to $A_{X}$ where $X$ stands for any
waveband, according to the standard interstellar extinction law from
Rieke \& Lebofsky (\cite{RL85}).  \ion{H}{i} column density should
thus be integrated along the line-of-sight for each stellar particle,
as Westera et al. (\cite{WSBG02}) did. Our SPH kernel is defined by
(Monaghan \& Lattanzio \cite{ML85}):
$$
W(r,h) = \frac{1}{\pi h^3}
\left\{
\begin{array}{ll}
        1-\frac{3 v^2}{2} + \frac{3 v^3}{4} & \mbox{if } 0\le v \le 1\\ \\
        \frac{(2-v)^3}{4} & \mbox{if } 1\le v \le 2 \\ \\
        0 & \mbox{otherwise}
\end{array}
\right.
$$
where $v=r/h$. The contribution of each particle to the density
vanishes for $r > 2h$. Thus, for each stellar particle, we should have
to draw up a list of gas particles of which the sphere of radius $2h$
is crossed by the line-of-sight. The gas column density is then the
sum of individual gas particle density weighted by the SPH kernel.

This technique is obviously more difficult and CPU time consuming than
with a grid. Therefore, we have computed extinction in a grid with the
same spatial resolution than our images and 11 slabs along the
line-of-sight from $z=-6.62$~kpc to $z=+6.62$~kpc. The slab limits are
spaced in $\log(|z|)$ in order to get a better resolution near the
equatorial plane where most of the gas is concentrated.  Each slab
absorbs the stellar luminosity behind it.  For each slab, the gas
density distribution is obtain by the convolution of particle
positions by the kernel.

\section{Global evolution over 1~Gyr}

The simulation has been stopped at 1~Gyr for the following reasons:
\begin{enumerate}
\item our simulation should be considered as a numerical
experiment. Most of the physical assumptions and recipes are simpler
than in most chemodynamic numerical codes that are used to make
simulations over several Gyr. This means that their validity might be
questionable on longer timescales (e.g. the use of Toomre $Q$
parameter).
\item the closed box assumption is unlikely valid over a longer
timescale. Gas infall, satellite harassment and/or merging, etc. are
some examples of external perturbation that invalidate the closed box
approximation.
\item since we do not add any dark halo (fixed or with live
particles), we neglect the interactions between the disc and the
halo. It is thus not reasonable to study the long term effect of the
disc evolution without any dark matter.
\end{enumerate}

\begin{figure*}
\centering
\vspace{5cm}
\caption{Snapshots of the evolution of the simulation from $t=0$ to
$t=1$~Gyr. From left to right are displayed the mass distribution and
B dust-free, B with extinction, K dust-free, and K with extinction
calibrated images. The field of view (20~kpc) is the same for each
frame. The range in mass is [1., 1.35\,10$^{5}$]~\densunit, displayed
in $\log$ units. B surface brightness ranges from 12.75 to
28.8~\magunit\ and K ranges from 10.9 to 26.2~\magunit. The particles
have been rotated so that the bar position is roughly horizontal}
\label{fig:contour}
\end{figure*}

\begin{figure*}
\centering
\vspace{5cm}
\caption{Evolution of the simulation from $t=0$ to $t=1$~Gyr. From
left to right are displayed the gas mass distribution, the \sfrdix\
and \sfrcent\ distributions and B$-$H colour map without and with dust
extinction. The field of view (20~kpc) is the same for each frame. For
the gaseous mass density, a lower cutoff has been apply for the sake
of clarity. The range in mass density is thus [10$^{-2}$,
2.5\,10$^{3}$]~\densunit, displayed in $\log$ units. \sfrdix\ and
\sfrcent\ ranges from $7.3\,10^{-4}$ (in the disc) to
365~\sfrdensunit\ (in the nucleus), also displayed in $\log$ units.
B$-$H ranges from $-0.57$ (black) to 6.14 (white). The particles have
been rotated so that the bar position is roughly horizontal}
\label{fig:contour2}
\end{figure*}

\subsection{Dynamic evolution}
\label{ssec:dynevol}

The initial disc quickly develops a typical strong bar and a spiral
structure (Fig.~\ref{fig:contour}) both in the stellar and the gaseous
components. The gravity torques due to the bar and spiral structure
drive the gas inwards and reorganise the mass distribution even for
the old stellar population; this gas inflow occurs in a rather short
timescale since the star formation rate peaks at $t=0.3$~Gyr
(cf. Fig.~\ref{fig:SFR}). As expected, star formation also occurs
along gaseous spiral arms.

After 1~Gyr, the total mass inside the central kpc has roughly
doubled. Indeed, a nuclear gas disc is formed from the accumulation of
gas in the centre, and new stars are actively formed there.  The first
cause of this mass inflow is the overall reorganisation of the mass
distribution under the influence of the stellar bar. Due to the
gravitational torques exerted on the gas by the stellar bar, the extra
mass in the form of gas and new stars amounts to
$3.5\,10^{9}$~M$_{\sun}$ at $t=1$~Gyr, which is only 30\% of the whole
additional mass. Indeed, the old stars population contributes to the
other 70\%.

The gas mass fraction inside the central kpc is very time
dependent. For instance, between $t=0.905$~Gyr and $t=1$~Gyr the mass
of gas is divided by a factor 4. This illustrates a property of the
gas inflow towards the central kpc region: the fueling is not
stationary but rather proceeds by burst.

Kinematical properties of the central kpc are described by Wozniak et
al.~(\cite{Wetal03}).

\begin{table}[ht]
\caption[ ]{Masses (in $10^{10}$ M$_{\sun}$) of the various population at $t=0$
and $t=1$~Gyr.}
\begin{flushleft}
\begin{tabular}{lrrrr}
\hline
\hline
Population      & \multicolumn{2}{c}{Whole simulation} & \multicolumn{2}{c}{Central kpc}\\
                & $t=0$  & $t=1$             & $t=0$  & $t=1$ \\
\hline
\noalign{\smallskip}
Old stars       & \multicolumn{2}{c}{$11.$}   & $1.04$  & $1.8$ \\
Gas             & $1.1$ & $0.53$       & $0.08$      & $0.0011$ \\
New stars       & $0$            & $0.57$       & $0$              & $0.43$ \\
Total mass      & \multicolumn{2}{c}{$12.1$}  & $1.12$  & $2.23$\\
\hline
\end{tabular}
\end{flushleft}
\label{tab:param}
\end{table}

\subsection{Star formation rate evolution}
\label{ssec:sfr}
\begin{figure}
\resizebox{\hsize}{!}{\includegraphics{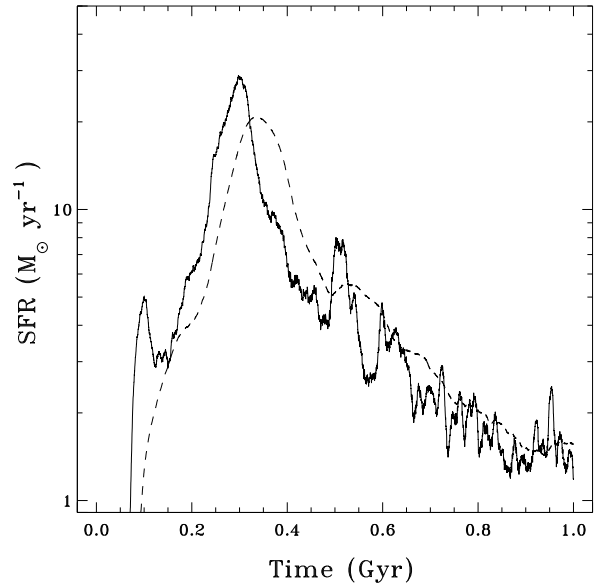}}
\caption{Global star formation rate (\sfrunit) versus the elapsed time
from the beginning of the simulation. Note that star formation is
inhibited during the first 60 Myr. Backward time average windows of
10~Myr (full line) and 100~Myr (dashed line) has been used}
\label{fig:SFR}
\end{figure}

The star formation rate (SFR) is displayed in Fig~\ref{fig:SFR}.  The
newly formed stellar mass is averaged backward over two time windows
of widths respectively 10~Myr (\sfrdix) and 100~Myr (\sfrcent).  Apart
from the first \sfrdix\ peak at $t=0.1$~Gyr which is due to the
progressive switch-on of the star formation in the code, the other
three \sfrdix\ maxima occurring around 0.3, 0.5 and 0.6~Gyr are
real. For $t \la 0.3$~Gyr, the bar and a bi-symmetric spiral structure
spontaneously form. The gravitational torques applied by the bar and
the spiral arms on the gas create several regions of very high gas
density in which star formation is ignited. \sfrdix\ reaches
30~\sfrunit. As shown in previous works (e.g. Martin \& Friedli
\cite{MF97}), the star formation is not homogeneously distributed over
the whole disc but is mainly concentrated along the bar major axis and
along the spiral arms (cf. Fig.~\ref{fig:contour2}). Indeed, $3/4$ of
the star formation occurs in the central kpc of the galaxy. SFR
density ranges from $7.3\,10^{-4}$ (in the disc) to 365~\sfrdensunit\
(in the nucleus). These values are in good agreement with data from
Kennicutt (\cite{K98}).

The next two peaks at $t \approx 0.5$ and 0.6~Gyr are the result of
the gas inflow towards the central region of the disc. Such inflow is
not continuous, but rather proceeds by burst.

When the 100~Myr averaging window is used, the SFR maximum is moved
forward by $\approx 50$~Myr.

\subsection{Luminosity and colours evolution}

\begin{figure}
\resizebox{\hsize}{!}{\includegraphics{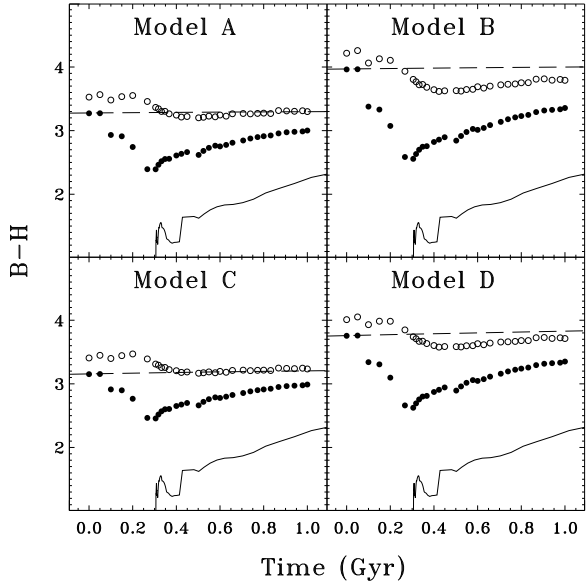}}
\caption{B$-$H colour evolution. Colours of the models without dust
extinction are represented by full points, with extinction by empty
points. The full line stands for the B$-$H index of an SSP with the
mean metallicity of gaseous star forming particles at $t=0.3$~Gyr. The
long-dashed curve is the B$-$H colour of the old population }
\label{fig:b-h}
\end{figure}

We plot in Fig.~\ref{fig:b-h} the evolution of the integrated B$-$H
colour index.  Our values are in good agreement with Scodeggio et
al. (\cite{Setal02}). Indeed, for their sample of spirals, B$-$H
values range between 1 and 4. For the sake of comparison, two SSP have
been overplotted. One is the SSP of the old population and thus
depends on the calibration model. The second is typical for a particle
born at the maximum SFR ($t=0.3$~Gyr). Its metallicity is
$Z_{\sun}/2$.

In absence of dust, the galaxy becomes bluer by $\approx 1$ magnitude
between $t=0$ and $t=0.25$~Gyr, then B$-$H slowly increases at a rate
similar to the SSP of the population born at the maximum SSP.  The
colour of the disc is thus dominated by the population born during the
most active phase of star formation around $t=0.3$~Gyr.

When dust extinction is taken into account, the amplitude of the B$-$H
evolution is strongly smoothed. B$-$H values remain close to the old
population SSP. For models B and D, a B$-$H minimum can still be
determined but it is shifted towards $t \approx 0.4-0.5$~Gyr. For
model C, which has a young initial population with a low metallicity,
B$-$H does not display any clear minimum. Indeed, the colour remains
roughly constant over 0.6~Gyr.

\begin{figure}[!h]
\resizebox{\hsize}{!}{\includegraphics{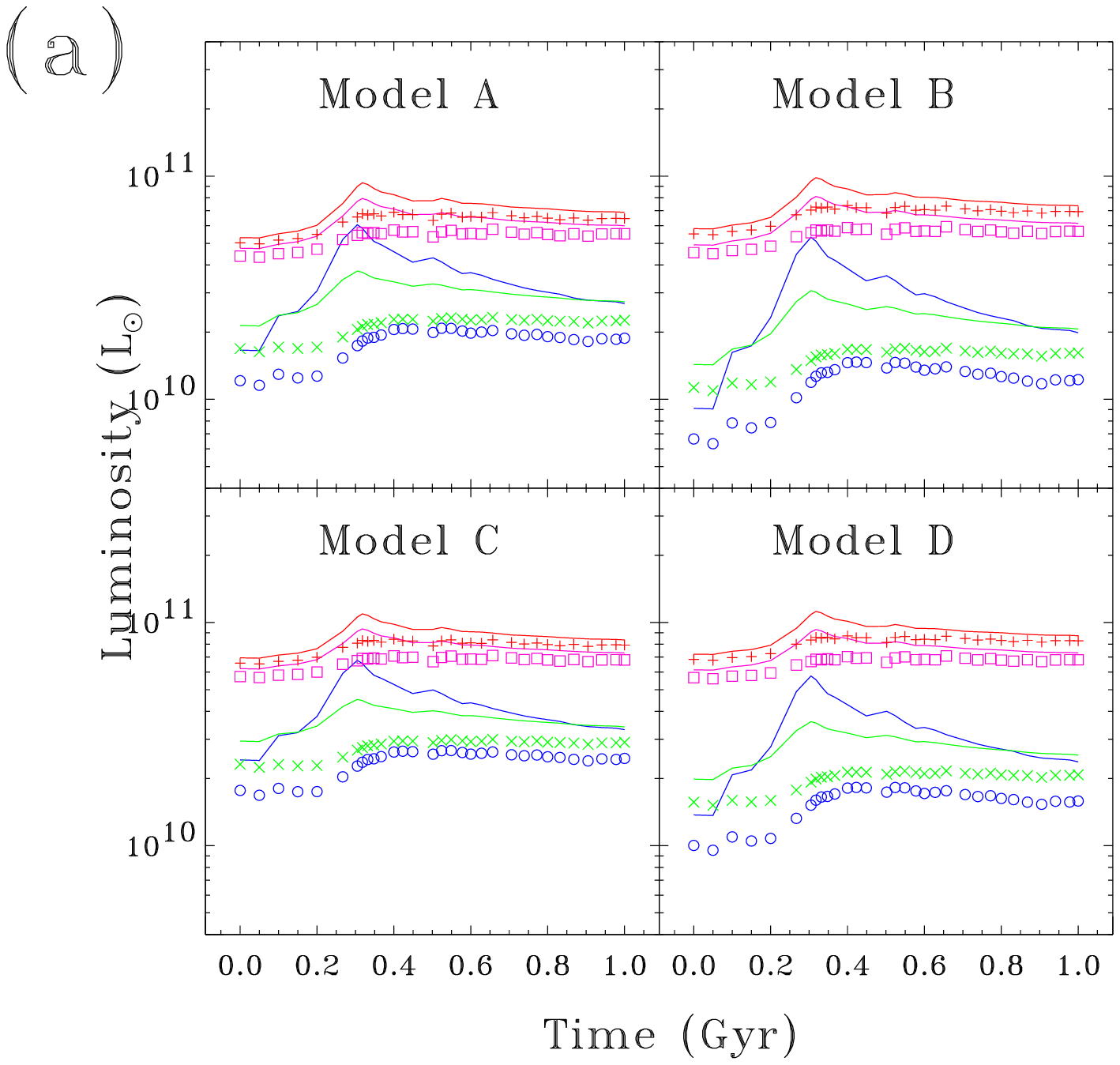}}
\resizebox{\hsize}{!}{\includegraphics{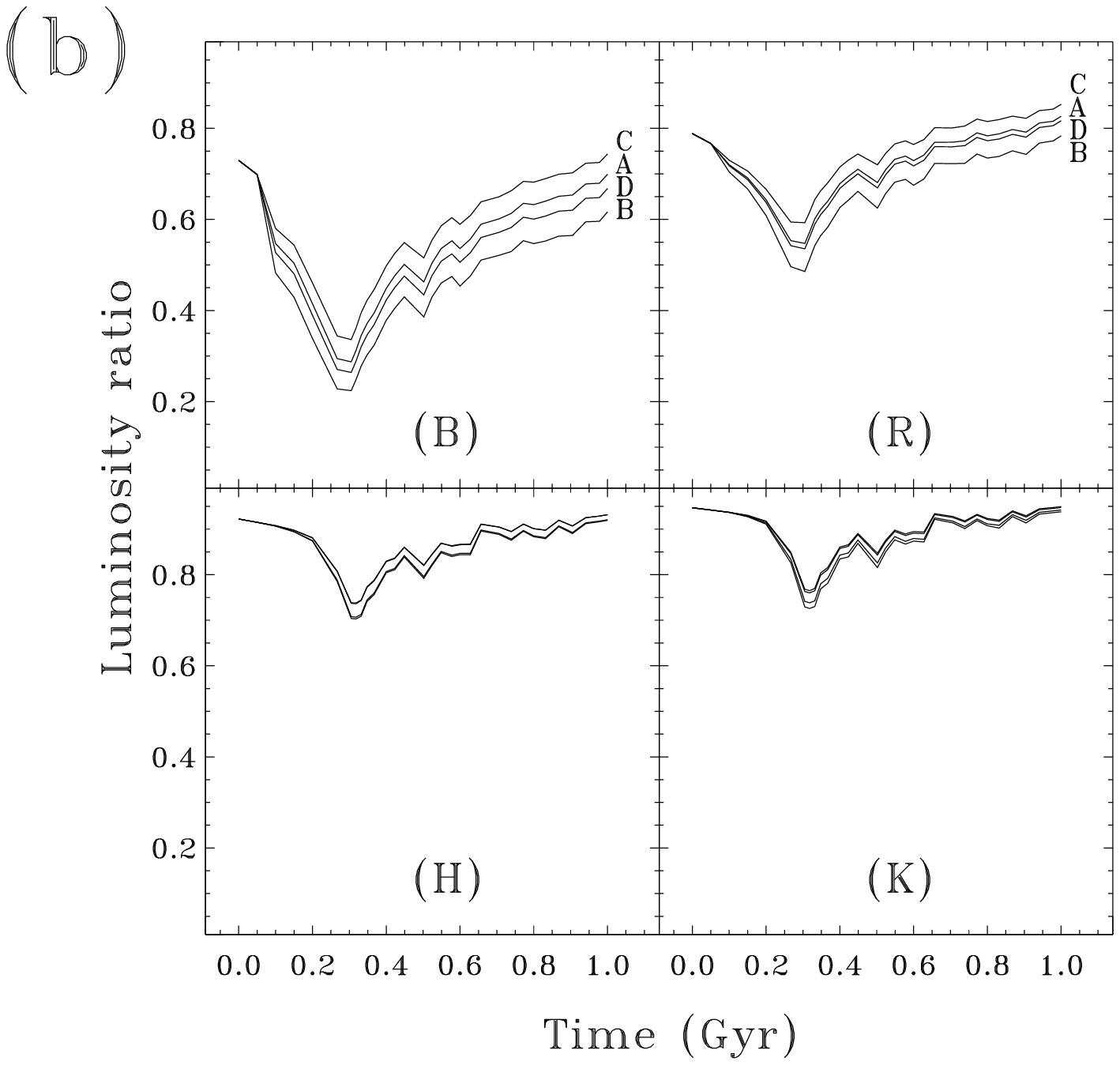}}
\caption{{\bf a)} Total luminosities for the four bands : B (blue), R
(green), H (magenta) and K (red). Lines represent the total luminosity
evolution of calibrated models without extinction (\ltot) while
symbols stand for absorbed luminosities (\ltotabs). Metallicity of the
old population increases from left ($Z=0.004$) to the right
($Z=0.02$), its age decreases from top (10.4~Gyr) to bottom (6.7~Gyr).
{\bf b)} Evolution of the ratio \ltotabs\ / \ltot\ in B (upper left
panel), R (upper right), H (lower left) and K (lower right). The
letters beside the lines refer to the calibration model}
\label{fig:ltot}
\end{figure}

Fig.~\ref{fig:ltot}a shows the total luminosity evolution for all the
models without (\ltot) and with dust extinction (\ltotabs).
Independently of the old population calibration, \ltot\ evolution
clearly shows two regimes. The broad peak centred around $t=0.3$~Gyr
obviously corresponds to the time when the star formation is the more
active. There is a small but significant lag between the \ltot\ peak
in the B and the infrared bands of $\approx 10$~Myr so that the B peak
is correlated to the \sfrdix\ maximum while the K peak occurs between
the \sfrdix\ and \sfrcent\ maximums. The maximum is followed by a slow
decrease which corresponds to the dimming of the stellar
population. However, the luminosity decrease is not exponential and is
rather slower than for a SSP. This is due to a rate of star formation
which remains high (above 2~\sfrunit) during the next 0.7~Gyr. This
continuous star formation partly compensates for the ageing of the
older stellar populations.

Hereafter we call luminosity contrast the ratio of the maximum value
of the luminosity to the luminosity at $t=0$. For our calibration
models, the latter corresponds to the luminosity of the old population
at the beginning of the simulation, whereas the maximum value is the
sum of luminosities of the old and of the new population.

The \ltot\ contrast depends on the old population calibration,
especially in the B band. For instance, the \ltot\ contrasts are
typically $\approx 1.7$ in K and between $\approx 3.8$ and $\approx
5.6$ in B according to the calibration model. However, these values
are not typical for a SSP since the contrast in the infrared is always
higher than in B. For instance, the contrasts amount to $\approx 10$
in K and $\approx 5$ in B for a SSP with $Z=0.004$.  This illustrates
that the old population predominates in mass and that the B band
luminosity of the old population is $\approx 2$ decade lower than the
K band luminosity.

As expected, when dust extinction is taken into account, the total
luminosities are fainter in all bands. However, since the amount of
gas as well as its spatial distribution evolve because of the combined
effects of the dynamic evolution and star formation, the ratio of
unobscured to obscured luminosity also evolves with time
(Fig.~\ref{fig:ltot}b).

It is also noteworthy that the luminosity peak at $t\approx 0.3$~Gyr
is smoothed out in R, H and K band. In B, \ltotabs\ still displays a
broad bump with a small maximum but shifted after $t > 0.4$~Gyr. This
delay is due to the fact that star formation with the highest SFR
occurs in the regions with the highest gas density. In these regions,
extinction is then the highest since we assume it is proportional to
the gas column density. Only between 1/4 and 1/3 of \ltot\ is able to
escape from the galaxy at $t=0.3$~Gyr
(cf. Fig.~\ref{fig:ltot}b). Thus, the regions which contain the
youngest population and thus contribute the most to the B band
luminosity are also the most obscured
(cf. Sect.~\ref{ssec:lumbar}). Since a fraction of the gas amount is
consumed during star formation and since gas and stars have different
kinematics, the gas density decreases during the evolution. This
enables the escape of the blue light after several tens of
Myr. Consequently, the \ltotabs\ maximum should not be interpreted as
a SFR maximum since it occurs $\sim 100$~Myr later on.

The high extinction of the youngest and brightest star forming regions
also explains that the ratio \ltotabs/\ltot\ is sensitive to the
parameters of the calibration model (cf. Fig.~\ref{fig:ltot}b)
although the dust extinction model is exactly the same for all
models. Indeed, since the B band luminosity of young particles is absorbed
by dust, the total B band luminosity is dominated by the contribution of
the old population. The differences between \ltot\ and \ltotabs\ are
more marked in B and R than in the infrared bands, since differences
in luminosity are more pronounced in B and R band than in infrared 
bands for the different old population. But the mean
extinction in K and H band is not negligible since the ratio
\ltotabs/\ltot\ amounts to 0.7 at $t=0.3$~Gyr.

\begin{figure}[!h]
\resizebox{\hsize}{!}{\includegraphics{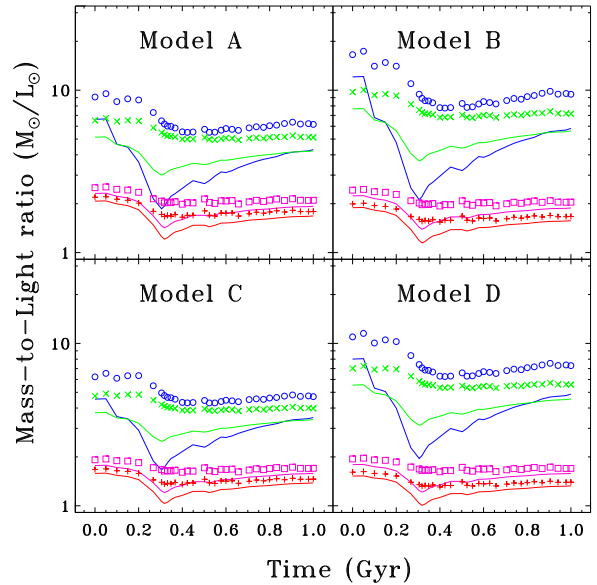}}
\caption{Total mass-to-light ratio evolution. Lines, symbols and
colours as for Fig.~\ref{fig:ltot}}
\label{fig:mltot}
\end{figure}

Stellar mass-to-light ratios are plotted in Fig.~\ref{fig:mltot}.
Although the total stellar mass increases with respect to time because
of star formation, the evolution of the total \ml\ is dominated by the
total luminosity evolution.  Values are in good agreement with mean
values of the RC3 catalogue for spiral type galaxies (Roberts \&
Haynes \cite{rh94}).  Some observations (e.g. Vallejo et al.
\cite{VBB02}) present lower values for \ml\ in K band, indicating the
presence of a very young population.

\subsection{Metallicity evolution}

\begin{figure}
\resizebox{\hsize}{!}{\includegraphics{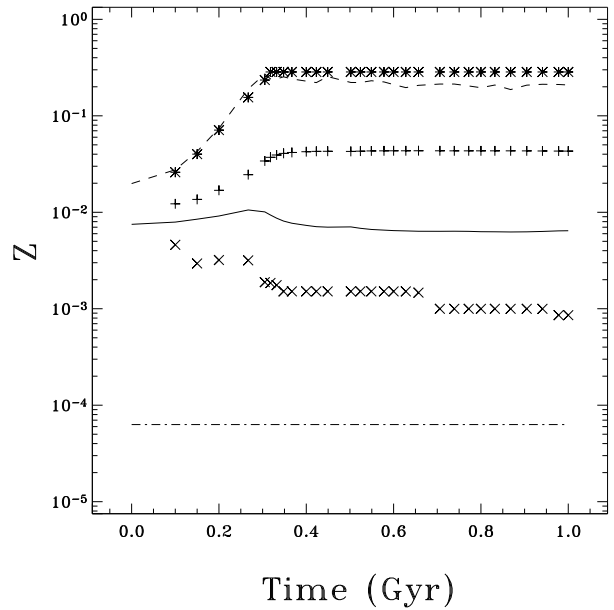}}
\caption{Metallicity evolution of gaseous particles (lines) and new
population particles (symbols). Mass-averaged (continuous line and
pluses), minimum (dot-dashed line and crosses) and maximum
metallicities (dashed line and stars) are drawn}
\label{fig:zevol}
\end{figure}

Since an initial gradient of metallicity has been imposed on the gas
particles at the beginning of the simulation
(cf.~Sect.~\ref{ssec:ci}), the metallicity of gas particles ranges
from $6 \times 10^{-5}$ (at the disc edge) to 0.02 (at the
centre). The mass-averaged metallicity (\barZ) is $7.5\times 10^{-3}$
(cf. Fig.~\ref{fig:zevol}). Then, \barZ\ and the maximum metallicity
increase until $t=0.3$~Gyr. During the same time, \barZ\ of the new
population also increases and is always higher than for the gas since
star formation is ignited in the central region of the disc. However,
during the next 0.7~Gyr, the mass of the gas significantly decreases
because of the strong star forming rate (\sfrdix $\approx
30$~\sfrunit). This implies that \barZ\ of the gas decreases while is
remains constant for the new population. Some gas particles with a
high metallicity are fully converted into new stars as their mass has
reach the lower threshold. This explains that the maximum metallicity
of the gas decreases.

At $t=1$~Gyr, some low metallicity gaseous particles still exist, but
because of the gas inflow towards the central region, their number
decreases. When star formation becomes active for such particles (for
instance when they encounter a spiral arm), they form new stars with a
low metallicity, so that the minimum metallicity for the new
population is a decreasing curve.

\section{Surface brightness evolution}

\subsection{Simulated images}

After individual particle calibration, simulated CCD images are
obtained summing particle luminosities into a 512$\times$512 pixels
grid. The field of view is 60~kpc which gives a spatial resolution of
$\approx 117$~pc, 1.3 times our smallest $N$-body grid resolution. We
thus produce one frame per waveband and per snapshot of the
simulation. Our results are obviously independent of the bar
position-angle with respect to the North or any other axis. We thus
decided to systematically rotate the positions of particles to align
the bar with the x-axis. An example is displayed in
Fig.~\ref{fig:contour} for model A.

\begin{figure}
\resizebox{\hsize}{!}{\includegraphics{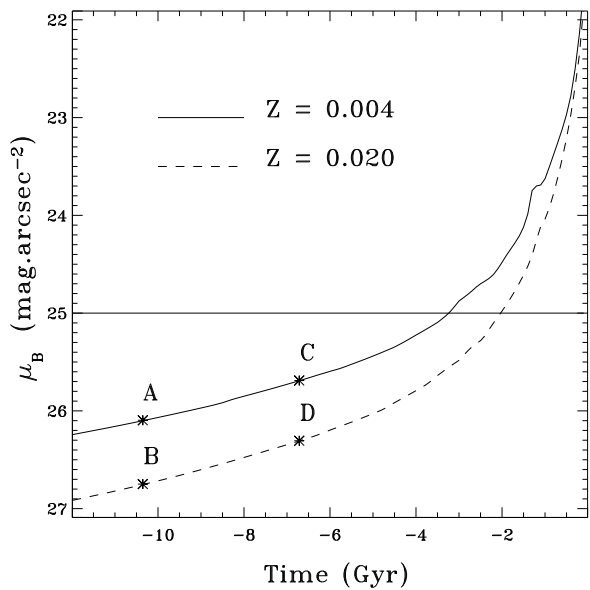}}
\caption{B band surface brightness of one particle of the old
population isolated in a cell of a CCD-like image at $t=0$. The lines
show the evolution of the surface brightness for the two metallicities
used for calibration. The origin of the time is the beginning of the
simulation ($t=0$) so as the age of the old population is
negative. The letters represent the model of calibration used for the
old population (cf. Table~\ref{tab:models}). The 25$^\mathrm{th}$
B band magnitude is plotted as an horizontal line for reference}
\label{fig:mu25}
\end{figure}

Due to a low number of particles in the external parts of the disc, we
have also build two times lower resolution images.  In
Fig.~\ref{fig:mu25} we show the magnitude of the faintest pixel (in
B~\magunit) that is reachable in a simulated image, depending on the
calibration model. This value corresponds to pixels which contains
only one particle. Because this only happens in the outskirts of the
disc, where star formation is inactive, these isolated particles
belong to the old population. This figure also show that our mass
resolution for the old population ($2\times 10^{5}$~M$_{\sun}$ per
particle) disables the possibility to study calibration models with an
initial age lower than 3.25~Gyr if we expect to determine an isophotal
diameter at the $25^\mathrm{th}$ magnitude in the B band.

To mimic real observations we should have to convolve our images with
a point spread function. However this last stage depends on the
telescope and site characteristics. It thus introduces a few free
parameters which cannot be constrained without any detailed comparisons
with real observations.

In Fig.~\ref{fig:contour2}, B$-$H colour maps are displayed for model
A, as well as the mass surface density of the gaseous component and
the \sfrdix\ and \sfrcent\ spatial distributions.  In absence of dust
extinction, B$-$H colour maps and \sfrcent\ distribution are similar,
which is no more the case when dust extinction is taken into
account. The extinction is not uniformly distributed in the bar
region. It is obviously concentrated in the region of highest gas
density and, therefore, in the region of high SFR.

\subsection{Ellipses fitting}

\begin{figure*}
\centering
\vspace{5cm}
\caption{For each panel: from top to bottom, surface brightness
($\mu$), ellipticity (e=1-b/a), position-angle (PA) and central shift
for each fitted ellipse, for the two bands (B plotted in blue and H in
red) and the mass distribution (in black). The mass density profiles
have been scaled so that the 30$^{th}$ magnitude corresponds to
1~\densunit. Results for dusty images are plotted in the form of dots}
\label{fig:profiles}
\end{figure*}

Each image has been analysed by fitting ellipses to the
isophotes. This technique is very efficient in retrieving embedded
structures (Wozniak et al. \cite{Wetal95}, Friedli et
al. \cite{Fetal96}). In general, it also gives reliable structural
parameters (Wozniak \& Pierce \cite{WP91}). The algorithm used in the
ellipse-fitting program is fully described in Jedrzejewski
(\cite{J87}). Thus, description of the procedure will be brief.

An isophotal contour of a given \smal\ is written as:
$$
I(\varphi)=c_0+\sum^{\infty}_{n=1}\, c_n\cos(n\varphi)+d_n\sin(n\varphi)
$$
If the trial ellipse exactly match the isophotal contour all the
coefficients but $c_0$ vanish. Then, $c_0$ is the mean intensity along
the contour. The departure from exact ellipticity can be quantified
with the Fourier components: errors in the position of the centre of
the ellipse give non zero values for $c_1$ and $d_1$ while errors in
ellipticity (defined by $e\!=\!1-b/a$ where $a$ and $b$ are
respectively the half-major and half-minor axis lengths) and/or
position-angle (PA) are revealed by $c_2$ and/or $d_2$ components.

Profiles of surface brightness, ellipticities and PA are
obtained by increasing the \smal\ $a$ by a factor $1.01$ between each
fit. We have chosen this small value to obtain a good resolution in
the inner region.

The mean intensity profiles are then converted to surface brightness
profiles. PA profiles follow the conventional notation. The North is
at the top, the East at the left and the angles are counted in the
anti-clockwise direction from 0\degr~ to 180\degr. The bulk of the
bar lies at PA$\approx 0$\degr. The range in radius for the figures
begins at 1 px, i.e. 117~pc; the upper bound corresponds roughly to a
surface brightness $\mu_{\rm B} \approx 10$~\magunit. We
show an example of these results in Fig.~\ref{fig:profiles} for the
model A at the same times than in Fig.~\ref{fig:contour}.

\subsection{Isophotal radii}

The surface brightness profiles obtained with ellipse fitting has been
used to determined the isophotal radius in each band. The surface
brightness levels at which isophotal radii are measured are 25 in B,
23.5 in R, 21.5 in K and 20.5 in H. For the H band, we follow Gavazzi
et al. (\cite{Getal00}) while Gavazzi et al. (\cite{Getal96}) used a
limiting magnitude of 21.5.

\begin{figure}
\resizebox{\hsize}{!}{\includegraphics{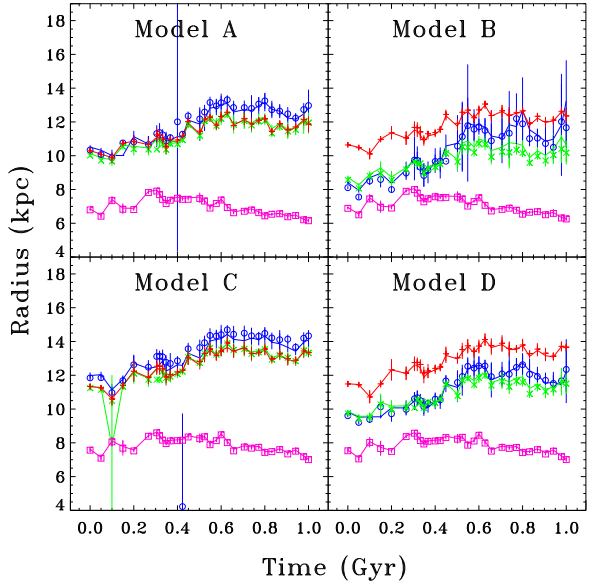}}
\caption{Evolution of the isophotal radii: \dB\ in B (blue), \dR\ in R
(green), \dH\ in H (magenta) and \dK\ in K (red). Symbols and lines as
for Fig.~\ref{fig:ltot}}
\label{fig:d25}
\end{figure}

\begin{figure}
\resizebox{\hsize}{!}{\includegraphics{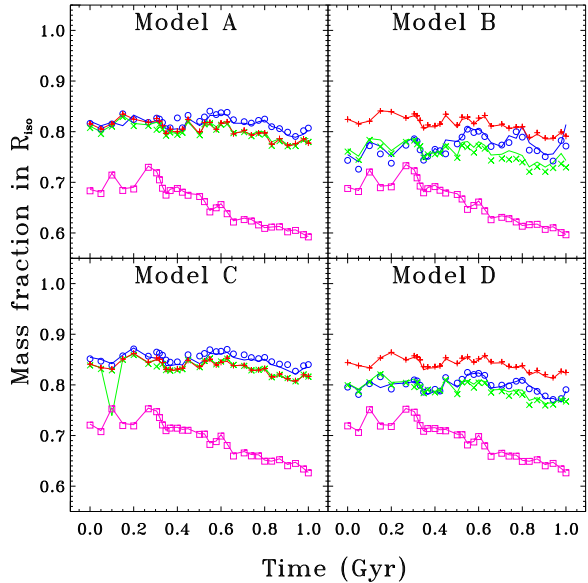}}
\caption{Evolution of the fraction of mass inside the isophotal radii:
\dB\ in B (blue), \dR\ in R (green), \dH\ in H (magenta) and \dK\ in K
(red). Symbols and lines as for Fig.~\ref{fig:ltot}}
\label{fig:massevol}
\end{figure}

\begin{figure}
\resizebox{\hsize}{!}{\includegraphics{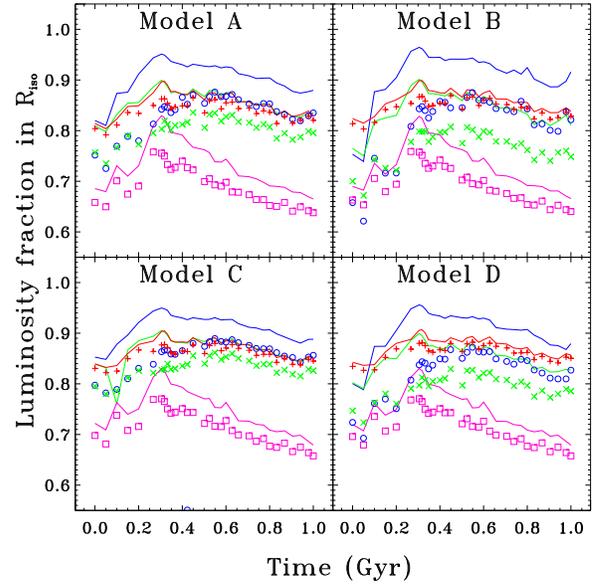}}
\caption{ Evolution of the fraction of luminosity integrated inside
the isophotal radii: \dB\ in B (blue), \dR\ in R (green), \dH\ in H
(magenta) and \dK\ in K (red). Symbols and lines as for
Fig.~\ref{fig:ltot}}
\label{fig:ld25}
\end{figure}

Fig.~\ref{fig:d25} displays the evolution of the isophotal radii for
all the models. Apart from the H band, the isophotal radii increase
with time during the first 0.6~Gyr and then remain constant or
slightly decrease until $t=1$~Gyr. In the H band, \dH\ clearly
decreases after the SFR reaches its maximum value ($t=0.3$~Gyr). The
isophotal radius of a stellar disc, dynamically stable and free from
star formation, should decrease with time since at a given radius the
surface brightness increases with time. However, the disc of our
simulation shows a strong dynamic evolution. We have shown in
Sect.~\ref{ssec:dynevol} that the mass is fully redistributed under
the influence of the bar. The mass inside the central kpc doubles
after 1~Gyr and 70\%\ of the additional mass come from the old
population. Thus, the evolution of the isophotal radii could be the
result of the combined effect of variations in luminosity due to star
formation and stellar evolution, and the dynamic evolution. To
disentangle the two effects, we plot in Fig.~\ref{fig:massevol} the
evolution of the fraction of mass inside the various isophotal radii
and, in Fig.~\ref{fig:ld25}, the evolution of the fraction of
luminosity inside isophotal radii.

The mass inside \dH\ clearly decreases. But the total H band
luminosity is rather constant or slowly increasing
(Fig.~\ref{fig:ltot}a), while the fraction of luminosity inside \dH\
drops from 80 to 65\%. Thus, the decrease of \dH\ only reflects the
dimming. It should be notice that \dH\ is not a good estimator of the
size of the galaxy since this radius only contains between 60 and
75\%\ of the total mass and less than 80\%\ of the total
luminosity. The case of \dK\ is similar although the mass decreases
slower.

The evolution of the mass inside \dB\ and \dR\ is more tricky since it
depends on the calibration of the old population. \dB\ and \dR\ are
between 10 and 20\%\ shorter when the metallicity of the old
population is solar while they are only 10\%\ shorter when the age of
the old population is greater. These differences are roughly constant.
Indeed, at these radii the old population dominates in mass and in
luminosity. So that the differences in isophotal radii between models
are directly linked to luminosity differences between old population.
Moreover these difference are roughly constant over 1~Gyr for a many
gigayears old population.

It is also obvious from Fig.~\ref{fig:d25} that dust extinction is
negligible at these radii since the gas density is very low. The
measurements of isophotal radii are thus robust.

Finally, Fig.~\ref{fig:ld25} also shows that the isophotal radii are
not a consistent estimator of the size of a galaxy. Indeed, depending
on the waveband and the delay with respect to the maximum SFR, these
radii delineate regions of various properties (luminosity, mass, etc.)
which cannot be compared between galaxies and, for the same object,
between different wavelength range.

\subsection{Bar characteristics}

\subsubsection{bar length and isophotal radii}

As displayed in Fig.~\ref{fig:contour} and Fig.~\ref{fig:contour2}, a
stellar bar forms, grows and becomes rounder during the simulation.
Thanks to our calibration, we are able to measure the bar length at
various wavelengths. Several criteria, defined in the literature, can
be used to determine the bar length. The results can then be compared
with dynamic radii.  This work will be published in a separate paper
(Michel--Dansac \& Wozniak \cite{pap2}).  Hereafter, we will use bar
lengths determined with a criterion shown to be in good agreement with
the corotation radius (namely \ctrois\ in Michel--Dansac \& Wozniak
\cite{pap2}). The bar lengths being similar in H and K band, we only
display the results obtained from the K band.

\begin{figure}
\resizebox{\hsize}{!}{\includegraphics{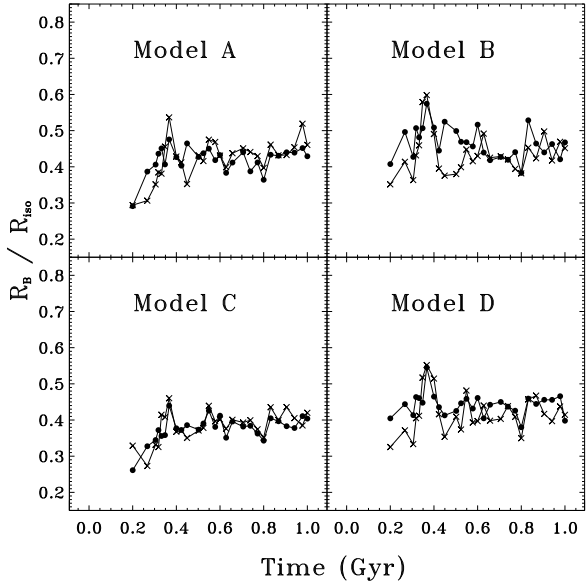}}
\caption{Evolution of the bar length to isophotal radius ratio: 
\dB\ in B (full circles), \dK\ in K (crosses)}
\label{fig:LbonD25}
\end{figure}

In Fig.~\ref{fig:LbonD25}, the evolution of the bar length to
isophotal radius ratio is plotted.  This ratio obviously increases
during the bar formation phase then remains roughly constant and
rather similar in both bands, for all the models apart from model B.
Between $t\approx 0.3$~Gyr and $t\approx 0.4$~Gyr there is a peak.

It should be notice that the difference between models are mostly due
to differences in the variation of the isophotal radius, since the
length of the bar is almost independent of the calibration model.

The values are comparable to observational data from Martin
(\cite{M95}). For a sample of isolated barred galaxies (Martinet \&
Friedli \cite{MtF97}), values of this ratio measured in B band range
between 0.04 and 0.47.  This ratio is insensitive to extinction effect
since dust extinction does not corrupt measurements neither of the
isophotal radii nor of the bar length.

\subsubsection{Luminosity of the bar region}
\label{ssec:lumbar}

We have computed the total luminosity inside a circle of radius equal
to the bar length. 

Comparison of our value with the literature is not
straightforward. Indeed, our definition of the bar luminosity includes
the bulge and the nucleus. Moreover, since the criterion used to
determine the end of the bar is a good tracer of the corotation
(cf. Michel--Dansac \& Wozniak \cite{pap2}), it gives upper limits of
the bar lengths. In the case of real observations, the bulge and
nucleus contributions are generally subtracted and bars end before
the corotation.  Thus, our values of the bar luminosity represent
upper limits.

\begin{figure}
\resizebox{\hsize}{!}{\includegraphics{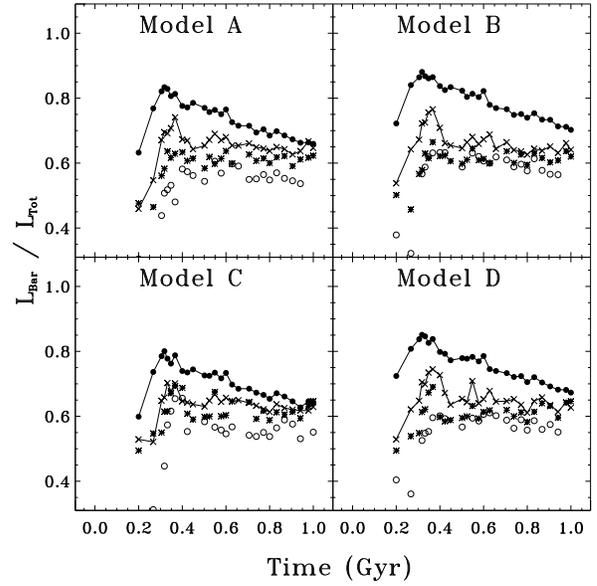}}
\caption{Evolution of the bar luminosity to total luminosity ratio in
B (full circles) and K (crosses). When extinction is taken into
account, the ratio is plotted as open circles in B and stars in K.}
\label{fig:lumbar}
\end{figure}

Evolution of the bar luminosity to total luminosity ratio is plotted
in Fig.~\ref{fig:lumbar} for the four models, in B and K band.  In B
band, the ratio increases rapidly, reaches its maximum during the SFR
peak, and then slowly decreases. A few weak bumps correspond to
secondary SFR peaks. According to the model, 80\%\ or more of the
total B band luminosity come from the bar region at the epoch of the
peak of SFR.  After 1~Gyr (SFR$<2$\sfrunit) more than two third of the
total blue luminosity still come from the bar region.  Extinction
effects in the bar region are important: at $t\approx 0.3$~Gyr less
than 20\%\ of the bar B band luminosity succeed to escape, and between
50 and 65\%\ at $t\approx 1$~Gyr according to the calibration models.

In the K band, the ratio increases rapidly, reaches a maximum
(70--75\%) during the SFR peak, and then rapidly (at $t \approx
0.4$~Gyr) reaches an approximatively constant value of about two third
of the total infrared luminosity.  Extinction effects are almost
negligible, just as model effects.

\subsubsection{Bar ellipticity}

\begin{figure}
\resizebox{\hsize}{!}{\includegraphics{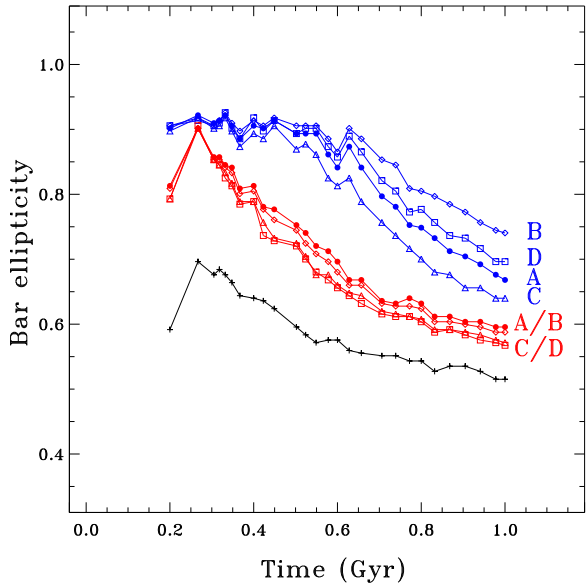}}
\caption{Bar maximum ellipticity for all the models. Results for the
mass distribution are displayed as black cross, red symbols for K band
images and blue symbols for B band images. The letters beside the
lines refer to the calibration model}
\label{fig:baremax}
\end{figure}

Maximum ellipticities reached in the bar region are displayed in
Fig.~\ref{fig:baremax}. Bars are significantly thinner in B than in K
band, except for $t = 0.27$~Gyr. In the B band, the bar is very thin
until $t \approx 0.55$~Gyr. The maximum ellipticity then decreases
with a slope which strongly depends on the calibration models. On the
contrary, the K band maximum ellipticity decreases immediately after
the SFR maximum. Hence, the difference between the two wavebands
increases until $t \approx 0.6$~Gyr, then it decreases with time but
never vanishes.

Two causes are simultaneously responsible of this effect. Firstly, the
spatial distribution of the new population is very elongated along the
bar because new stars born in the gas flow along the bar which is
narrower than the stellar bar (cf. Fig.~\ref{fig:contour2}). Then,
secular dynamic diffusion is responsible for progressively making it
rounder. This explains the high ellipticity reached in the bar when
SFR is high and, only in part, the subsequent decrease.  Secondly, the
luminosity ratio between the new and the old populations is wavelength
dependent, begin higher in B than in K band. Moreover, it roughly
follows the SFR variations. The colour dependence slowly becomes
blurred as SFR decreases. Hence, after the SFR peak, morphology
becomes gradually dominated by the luminosity of the old population
that has a rounder spatial distribution. This happens earlier in K
band. In B band, the maximum ellipticity decreases faster for more
luminous old population (model C). This effect is weaker in K band and
is linked to the age rather than metallicity of the old population.

The maximum ellipticity of the mass distribution reaches its maximum
at $t = 0.27~Gyr$, then slowly decreases. The formation of the bar and
its thickening occurs in the old stellar population which dominates
the mass. However, the mass distribution is always rounder than the
luminosity one. Since difference decreases with SFR, it could be
anticipated that isophotes and mass isodensities have the same shape
if the SFR inside the bar vanishes.  Thus, the surface brightness
distribution, even in the near infrared, could be a good tracer of
mass only if SFR remains low during a long time. Thus, mass modelling
that used surface brightness distribution might give biased results if
star formation is active, even at a moderate rate.

Dust extinction has obviously a more disturbing effect for B isophotes
than K ones. Since extinction is proportional to \ion{H}{i} column
densities, it decreases with time following gas consumption by star
formation.  Ellipticity profiles are fully different and very noisy
when extinction is high (cf. Fig.~\ref{fig:profiles}), making maximum
ellipticity determination difficult. However, maximum ellipticity
values do not significantly differ from the dust-free case\footnote{an
error is slipped into Fig.~2 of Michel--Dansac \& Wozniak
(\cite{kiel}). On the left panel, we compare maximum ellipticity in B
band with and without extinction, whereas on the right panel, we
compare maximum ellipticity in B and H band.}.  Differences on B band
ellipticity profile are still present at $t=1$~Gyr but appear mainly
in the central kpc.

\section{Conclusions}
 
We have carried out a photometric calibration of a self-consistent
N-body simulation of an isolated disc galaxy, including stars, gas and
star formation. Dust extinction has been taken into account. We have
simultaneously study the dynamic and photometric evolution in various
wavebands (B, R, H and K). We are able to reproduce observational
properties such as \ml\ and B$-$H colour indices. A morphological and
photometric analysis have also been performed on simulated images in
various wavebands, using the same tools as for real observations.

Our main results can be summarised as follows:
\begin{enumerate}

\item a dynamically self-consistent SFR peak produces a large increase
in total luminosity in all wavebands whereas the stellar mass of the
new population is much lower than the underlying old stellar
mass. However, luminosity peaks can be almost completely obscured by
dust extinction, since the young stellar population initially seats in
gas-rich regions where extinction is maximum. This effect is
particularly important on B$-$H colour maps since regions of highest
SFR could become unobservable. It could lead to underestimate SFR from
photometric or colour measurements.

\item we show how isophotal radii evolve with respect to mass
redistribution during the dynamic evolution but also with respect to
stellar evolution. Hence, the dynamic properties of the area enclosed
by any isophotal radius depends on the waveband and on the star formation
activity. Extinction effects do not affect the determination of
isophotal radii, even in B band. It is also noteworthy that the
luminosity integrated inside the isophotal radius at 20.5~\magunit\ in
the H band is not a good tracer of mass.

\item the luminosity integrated into the bar length reaches
$\approx 80$\%\ of the total B band luminosity at the SFR maximum in absence
of dust, and $\ga 60$\% when SFR is low, whereas the bar length remains
smaller than or equal to half the isophotal radius.

\item when SFR is high ($\approx 30$~\sfrunit) less than 20\%\ of the
B band luminosity of the bar region succeed to escape in the presence
of dust because most star formation occurs in gas-rich regions where
extinction is the most efficient.

\item as long as star formation is active, mass isodensities of the
bar region are rounder than isophotes. Thus, the surface brightness
distribution, even in H band, is a good tracer of mass only when
SFR is below 1~\sfrunit. Mass modelling that used surface brightness
distribution might give biased results if star formation is active,
even at a moderate rate.

\end{enumerate}

The use of other morphological classification tools
(e.g. concentration and asymmetry, Abraham et al. \cite{Aetal96}) will
be will be reported in a forthcoming paper.

\begin{acknowledgements}
Our computations were partly performed on the Fujitsu NEC SX-5 hosted
by IDRIS/CNRS. We are grateful to P. Berzcik and an anonymous referee
for fruitful comments which increase the legibility of our paper.
\end{acknowledgements}
 


\begin{thebibliography}{}
\bibitem[2003]{ANSE03} Abadi M.G., Navarro J.F., Steinmetz M., Eke
  V.R., 2003, \apj\ 591, 499
\bibitem[1996]{Aetal96} Abraham R.G., Tanvir N.R., Santiago B.X., et al.,
  1996, \mnras\ 279, L47
\bibitem[1989]{BH89} Boehringer H., Hensler G., 1989, \aap 215, 147 
\bibitem[2002]{BC02} Bournaud F., Combes F., 2002, \aap\ 392, 83
\bibitem[1993]{BC93} Bruzual G., Charlot S., 1993, \apj\ 405, 538
\bibitem[1981]{CS81} Combes F., Sanders R.H., 1981, \aap\ 96, 164
\bibitem[1993]{FB93} Friedli D., Benz W., 1993, \aap\ 268, 65
\bibitem[1995]{FB95} Friedli D., Benz W., 1995, \aap\ 301, 649
\bibitem[1994]{FBK94} Friedli D., Benz W., Kennicutt R., 1994, \apjl\
  430, L105
\bibitem[1993]{FM93} Friedli D., Martinet L., 1993, \aap\ 277, 27
\bibitem[1996]{Fetal96} Friedli D., Wozniak H., Rieke M., Martinet L.,
Bratschi P, 1996, \aaps\ 118, 461
\bibitem[1997]{GI97} Gerritsen J.P.E., Icke V., 1997, \aap\ 325, 972
\bibitem[1996]{Getal96} Gavazzi G., Pierini D., Boselli A., Tuffs R.,
1996, \aaps\ 120, 489
\bibitem[2000]{Getal00} Gavazzi G., Franzetti P., Scodeggio M.,
Boselli A., Pierini D., 2000, \aap\ 361, 863
\bibitem[1994]{HS94} Heller C.H., Shlosman I., 1994, \apj\ 424, 84
\bibitem[1978]{H78} Hohl F., 1978, \apj\ 83, 768
\bibitem[1987]{J87} Jedrzejewski R.I., 1987, \mnras\ 226, 747
\bibitem[1991]{KG91} Katz N., Gunn J.E., 1991, \apj\ 377, 365
\bibitem[1992]{K92} Katz N., 1992, \apj\ 391, 502
\bibitem[2001]{K01} Kawata D., 2001 \apj\ 558, 598
\bibitem[1989]{K89} Kennicutt R.C., 1989, \apj 344, 685 
\bibitem[1990]{K90} Kennicutt R.C., 1990, ASSL Vol.~161: The
  Interstellar Medium in Galaxies, 405 
\bibitem[1998]{K98} Kennicutt R.C., 1998, \araa\ 36, 189
\bibitem[1992]{Letal92} Leitherer C., Robert C., Drissen L., 1992, \apj\ 401, 596 
\bibitem[1996]{MH96} Mihos J.C., Hernquist L., 1996, \apj\ 464, 641 
\bibitem[1992]{M92} Maeder A., 1992, \aap\ 264, 105
\bibitem[1995]{M95} Martin P., 1995, \aj\ 109, 2428
\bibitem[1997]{MF97} Martin P., Friedli D., 1997, \aap\ 326, 449
\bibitem[1997]{MtF97} Martinet L., Friedli D., 1997, \aap\ 323, 363
\bibitem[2003]{kiel} Michel--Dansac L., Wozniak H., 2003, \apss\ 284, 925
\bibitem[2004]{pap2} Michel--Dansac L., Wozniak H., 2004, submitted
\bibitem[1985]{ML85} Monaghan J.J., Lattanzio J.C., 1985, \aap\ 149, 135
\bibitem[1996]{NSH96} Norman C.A., Sellwood J.A., Hasan H., 1996,
  \apj\ 462, 114
\bibitem[1993]{PF93} Pfenniger D., Friedli D., 1993, \aap\ 270, 561
\bibitem[1990]{PN90} Pfenniger D., Norman C., 1990, \apj\ 363, 391
\bibitem[1985]{RL85} Rieke G.H., Lebofsky M.J., 1985, \apj\ 288, 618 
\bibitem[1994]{rh94} Roberts M.S., Haynes M.P., 1994, \araa\ 32, 115
\bibitem[2003]{SG03} Samland M., Gerhard O., 2003, \aap\ 399, 961
\bibitem[1959]{S59} Schmidt M., 1959, \apj\ 129, 243 
\bibitem[2002]{Setal02} Scodeggio M., Gavazzi G., Franzetti P., et al.,
  2002, \aap\ 384, 812
\bibitem[1993]{SN93} Shlosman I., Noguchi M., 1993, \apj\ 414, 474
\bibitem[1995]{SM95} Steinmetz M., Muller E., 1995, \mnras\ 276, 549
\bibitem[1964]{T64} Toomre A., 1964, \apj\ 139, 1217
\bibitem[2002]{VBB02} Vallejo O., Braine J., Baudry A., 2002, \aap\ 387, 429
\bibitem[2002]{WSBG02} Westera P., Samland M., Buser R., Gerhard O.E.,
2002, \aap\ 389, 761
\bibitem[1991]{WP91} Wozniak H., Pierce, M.J., 1991, \aaps\ 88, 325
\bibitem[2003]{Wetal03} Wozniak H., Combes F., Emsellem E., Friedli
D., 2003, \aap\ 409, 469
\bibitem[1995]{Wetal95} Wozniak H., Friedli D., Martinet L., et al.,
1995, \aaps\ 111, 115
\end{thebibliography}
\end{document}